\documentclass{aa}
\usepackage{graphicx}
\usepackage{txfonts}
\usepackage{natbib}
\usepackage{color} 
\definecolor{highlight}{rgb}{0,0,0}
\newcommand{\highlight}[1]{{{\color{highlight} #1}}}

\title{Near-infrared interferometric observation of the Herbig Ae star HD144432 with VLTI/AMBER
  \thanks{
    Based on observations made with ESO telescopes at Paranal Observatory under
    program {\color{black}ID 083.D-0224(C) and 085.C-0126(A)}.
  }
}

\author{L. Chen\inst{1}
\and A. Kreplin\inst{1}
  \thanks{Member of the International Max Planck Research School (IMPRS)
    for Astronomy and Astrophysics at the Universities of Bonn and Cologne.}
\and Y. Wang\inst{1}
\and G. Weigelt\inst{1}
\and K.-H. Hofmann\inst{1}
\and S. Kraus\inst{2}
\and \\D. Schertl\inst{1}
\and S. Lagarde\inst{3} 
\and A. Natta\inst{4} 
\and R. Petrov\inst{3} 
\and S. Robbe-Dubois\inst{3} 
\and E. Tatulli\inst{5,6} 
}

\institute{
Max-Planck-Institut f\"ur Radioastronomie, Auf dem H\"{u}gel 69, 53121 Bonn, Germany
\\email: lchen@mpifr-bonn.mpg.de
\and Department of Astronomy, University of Michigan, 500 Church St., Ann Arbor, MI 48109-1090, USA
\and Laboratoire Lagrange, UMR7293, Universit\'e de Nice Sophia-Antipolis,  
CNRS, Observatoire de la C\^ote d'Azur, 06300 Nice, France
\and INAF -- Osservatorio Astrofisico di Arcetri, Largo E. Fermi 5, 50125 Firenze, Italy
\and UJF-Grenoble 1/CNRS-INSU, Institut de Plan\'etologie et d'Astrophysique de Grenoble (IPAG) UMR 5274,
	Grenoble, F-38041, France
\and Inter-University Centre for Astronomy and Astrophysics (IUCAA),
Post Bag 4, Ganeshkhind, Pune 411007, India
}

\newcommand{\TGM}{temperature-gradient model}
\newcommand{\GM}{geometric model}

\newcommand{\degree}{{}^\circ}
\newcommand{\mas}{{\mathrm{mas}}}
\newcommand{\AU}{{\mathrm{AU}}}
\newcommand{\pc}{{\mathrm{pc}}}
\newcommand{\mum}{{\mu\mathrm{m}}}
\newcommand{\enDash}{{\textrm{--}}}

\begin{document}

\abstract
{} 
{
We study the sub-AU-scale circumstellar environment of
the Herbig Ae star HD144432 with near-infrared (NIR) VLTI/AMBER observations
to investigate the structure of its inner dust disk.
\rm
} 
{
The interferometric observations were carried out with the AMBER instrument in the $H$ and $K$ band.
We interpret
the measured  $H$- and $K$-band visibilities,
the near- and mid-infrared visibilities from
the literature,
and the SED of HD144432
by using geometric ring models and ring-shaped temperature-gradient disk models
with power-law temperature distributions.
} 
{
We derived a $K$-band ring-fit radius of $0.17\pm0.01~\AU$ and an $H$-band radius of $0.18\pm0.01~\AU$
(for a distance of $145~\pc$).
This measured $K$-band radius of ${\sim}0.17~\AU$ lies in the range between
the dust sublimation radius of ${\sim}0.13~\AU$
(predicted for a dust sublimation temperature of $1500~\mathrm{K}$ and gray dust)
and the prediction of models including backwarming (${\sim}0.27~\AU$).
We found that an additional extended halo component is required in both the geometric and {\TGM}ing.
In the best {\TGM}, the disk consists of two components.
The inner part of the disk is a thin ring with an inner radius of ${\sim}0.21~\AU$,
a temperature of ${\sim}1600~\mathrm{K}$,
and a ring thickness ${\sim}0.02~\AU$.
The outer part extends from ${\sim}1~\AU$ to ${\sim}10~\AU$
with an inner temperature of ${\sim}400~\mathrm{K}$.
We find that the disk is nearly face-on with an inclination angle of $<28\degree$.
} 
{
Our {\TGM}ing suggests that the NIR excess is dominated by emission
from a narrow, bright rim located at the dust sublimation radius,
while an extended halo component contributes ${\sim}6\%$ to the total flux at $2~\mum$.
The MIR model emission has a 
two-component structure
with ${\sim}20\%$ flux from the inner ring
and the rest from the outer part.
This two-component structure suggests a disk gap, which is possibly caused by the shadow of a puffed-up inner rim.
} 

\keywords{stars: individual: HD144432
  - stars: pre-main sequence
  - circumstellar matter
  - techniques: interferometric
  - planetary systems: planetary disks
  - accretion: accretion disks
  }

\maketitle

\section{Introduction}

\object{HD144432} (He3-1141) is an isolated Herbig Ae (HAE) star
with spectral type A9/F0 \citep{The1994, Sylvester1996}
located at ${\sim}145~\pc$ (\citealp{Perez2004}; see also the discussion in Sect. 4).
Since it is not closely associated with molecular cloud material \citep{Malfait1998},
its IR excess can be attributed to circumstellar (CS) material.
By analyzing its spectral features, and considering its lack of photometric variability,
\citet{Meeus1998} concluded that the central star of HD144432 is directly seen, without CS material in the line of sight.
Thus, they suggested that the CS material is confined to a face-on disk, instead of having a spherical geometry.
HD144432 belongs to the group II objects in the classification scheme by \citet{Meeus2001},
i.e., it has a flat IR spectrum and a weaker MIR excess than the group I objects.
A plausible explanation for a group II SED is that
the dust disk is shadowed by its own puffed-up inner rim
\citep{Natta2001,Dullemond2001,Dullemond2002,Dominik2003,Dullemond2004a}.
The polycyclic-aromatic-hydrocarbon emission features in HD144432 are weak \citep{Acke2004, Keller2008},
which supports the self-shadowed-disk interpretation.

Previous interferometric observations of HD144432 in the $H$ and $K$ bands
using IOTA and the Keck Interferometer (KI)
were reported by \citet[][hereafter M05]{Monnier2005},
\citet[][hereafter M06]{Monnier2006},
and \citet{Eisner2009}.
Spectrally dispersed interferometric $K$-band observations by \citet{Eisner2009}
show that the size of the emission region increases with wavelength.
Their modeling of the data suggests
a dust disk with an inner radius of $0.25\textrm{--}0.3~\AU$
(at $145~\mathrm{pc}$)
and an inner temperature of $1000\textrm{--}1200~\mathrm{K}$,
a gas disk between the star and the dust,
and  a $\mathrm{Br}\gamma$ line originating from a more compact region.
In the mid-infrared (MIR), \citet{Leinert2004} measured the half-light radius of the disk at $12.5~\mu\mathrm{m}$
to $14~\mas$ (${\sim}2~\AU$ at $145~\mathrm{pc}$) using VLTI/MIDI.
With Keck segment-tilting,
\citet{Monnier2009} measured the Gaussian FWHM of HD144432
to $39\pm5~\mas$ ($5.6\pm0.7~\AU$ at $145~\mathrm{pc}$) at $10.7~\mu\mathrm{m}$.

\citet{Perez2004} found that the object is a binary, with a K-type T Tauri star companion at a separation of $1.4''$,
and estimated the age of the two stars to $1\textrm{--}3~\mathrm{Myr}$.
\citet{Carmona2007} confirmed that the primary and the companion are physically associated, but estimated the age to $8~\mathrm {Myr}$.

In this paper, we present $H$- and $K$-band VLTI/AMBER observations of HD144432.
In Sect. 2, we summarize the observations and the data reduction.
The modeling is presented in Sect. 3.
We discuss the modeling results in Sect. 4 and present the summary and conclusion in Sect. 5.


\section{Observation and data reduction}
AMBER is the NIR beam combiner instrument of the Very Large Telescope Interferometer (VLTI) and
records spectrally dispersed three-beam interferograms, capable of measuring both visibilities and closure phases (CPs) \citep{Petrov2007}.
HD144432 was observed in the low spectral resolution mode (R = 35) in the $H$ and $K$ bands
on 2009 Apr 18 and 2010 Apr 18 with VLTI/AMBER
using the linear baseline configuration E0-G0-H0 and the triangle configuration D0-H0-G1, respectively (see Table 1).
The visibilities and CPs are derived using the Pixel-to-Visibilty-Matrix (P2VM) algorithm 
of the data reduction package amdlib 3.0%
\footnote{available at: http://www.jmmc.fr/data\_processing\_amber.htm}.
In Fig. 1, we show the $H$- and $K$-band visibilities as a function of projected baseline length.
The $1.4''$ binary companion has no influence on the visibility measurements of the primary star,
since the field-of-view of the AMBER is only $0.25''$.
\highlight{
The extracted CPs are shown in Fig. 2 and have wavelength-averaged values
of $1.5\pm1.7\degree$ (2009), $0.3\pm5.0\degree$ (2010a) and, $-3.1\pm5.9\degree$ (2010b). 
In Fig. 3a (upper three rows), we present the derived wavelength-dependent AMBER visibilities
of HD144432 in the $H$ and $K$ bands.
}
For data processing, we selected 30\% of the frames with the highest fringe signal-noise-ratio \citep{Tatulli2007}
of each target and calibrator data set.
From the 2009 and 2010a data (see Table 1), we extracted both $H$- and $K$-band visibilities and CPs.
From the 2010b data, only $K$-band visibilities and CPs were extracted due to low fringe SNR of the $H$-band data.
We used the method of OPD histogram equalization \citep{Kreplin2011}
to reduce the influence of atmospheric optical path differences (OPDs) on the calibrated visibilities. 
In the calibration process, we used in both nights the calibrator star HD142669
with an uniform disk diameter of $d_{\mathrm{UD}} = 0.27 \pm 0.05$ mas%
\footnote{Taken from the Catalogue of Stellar Diameters (CADARS) \citep{Pasinetti-Fracassini2001}}.
\begin{table*}
\caption{Observation log of our VLTI/AMBER observation of HD144432.}
\label{obslog}
\centering
\begin{tabular}{cccccccc}
\hline
\hline
Data set  & Night  & $t_{\rm obs}$ & telescope configuration & $B_{\rm p}$ & PA   & Seeing & DIT\tablefootmark{a}\\ 
 &             & (UTC)         &      & (m)         & ($^\circ$) & ($\arcsec$) & (ms) \\
\hline
2009   & 2009-04-18 & 04:02:36 & E0-G0-H0 & 13.4 / 26.7 / 40.1 & 41.7                & 0.72 & 200   \\
2010a  & 2010-04-18 & 07:57:31 & D0-H0-G1 & 62.5 / 71.2 / 71.3 & 76.9 / 140.6 / 12.8 & 0.75 & 200 \\
2010b  & 2010-04-18 & 10:20:02 & D0-H0-G1 & 46.2 / 67.0 / 68.9 & 92.4 / 164.8 / 24.5 & 0.59 & 200 \\
  \hline
\end{tabular}
\highlight{
  \tablefoot{
  \\ \tablefoottext{a}{Detector integration time.}
  }
}
\end{table*}

\section{Modeling}

\begin{figure}
\center
\includegraphics[scale=0.35,angle=-90]{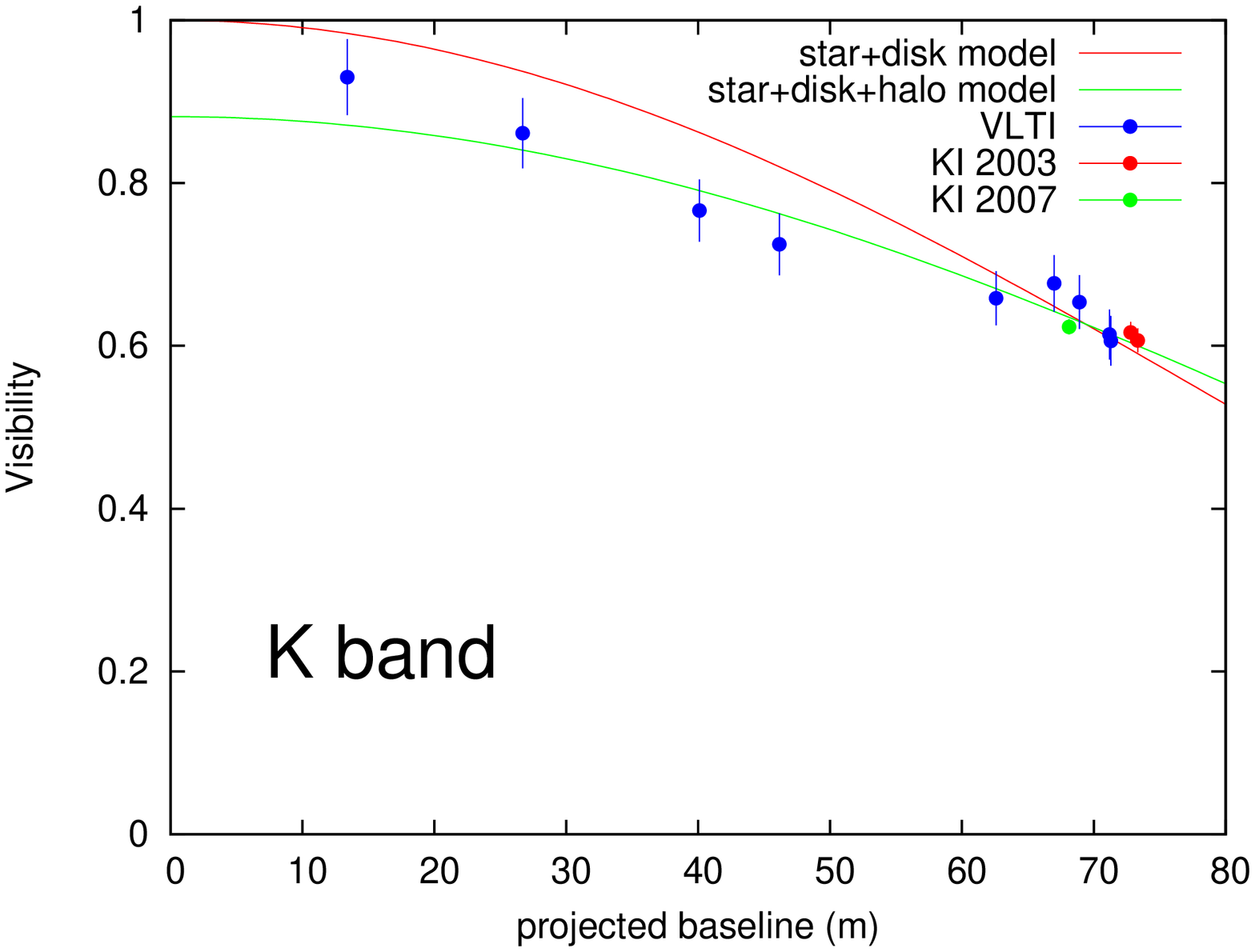} \\
\includegraphics[scale=0.35,angle=-90]{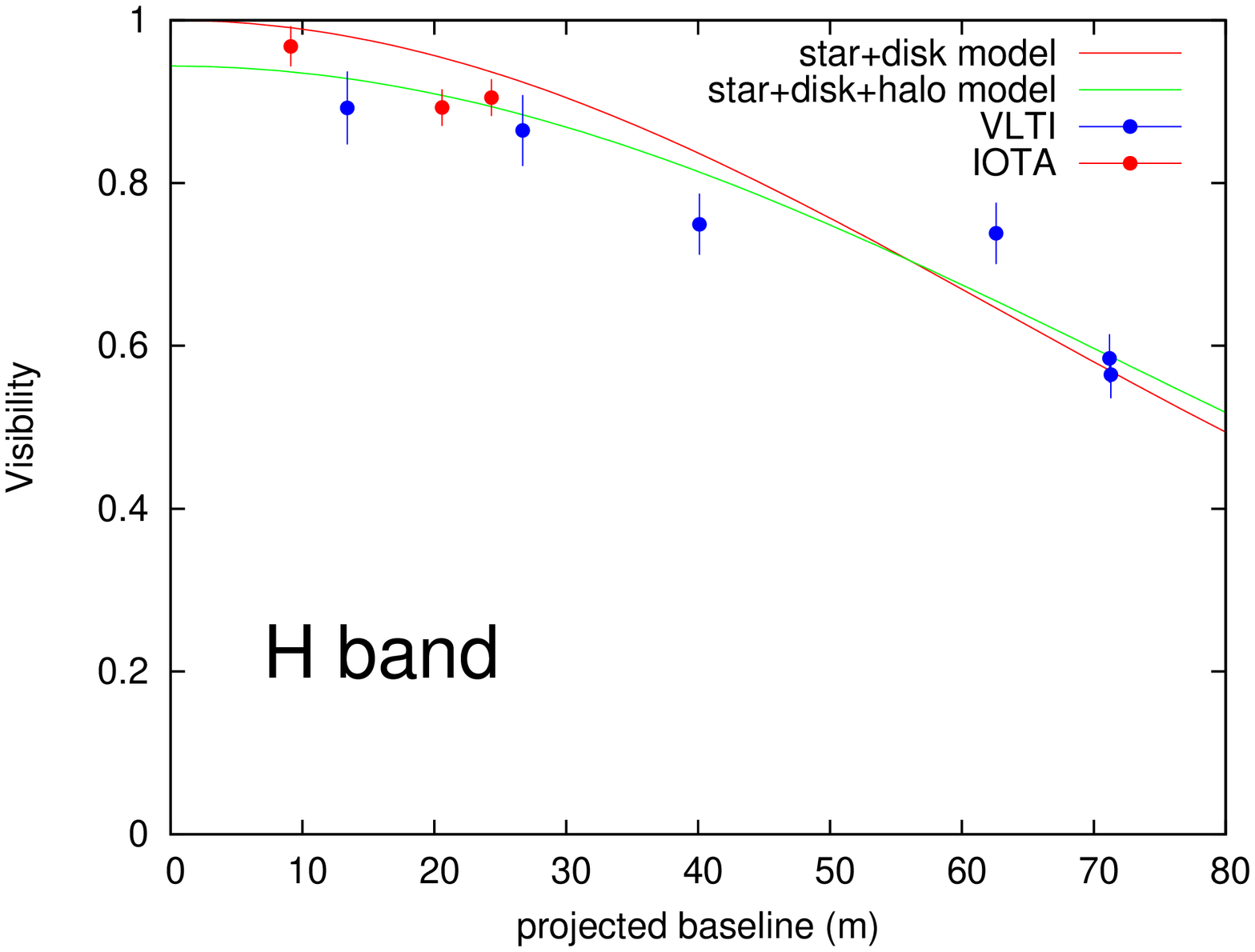}
\caption{
  Band-averaged visibilities as function of baseline length.
  The dots with errorbars are the observations
  (blue dots: our VLTI data;
  red and green dots: IOTA and KI data taken from \citealt{Monnier2005,Monnier2006} and \citealt{Eisner2009}%
  ).
  The lines are best-fit {\GM}s
  (inclination $0\degree$, red: star-disk model; green: star-disk-halo model).
  Top panel: $K$ band.
  Bottom panel: $H$ band.
  Model parameters are listed in Table \ref{tab:TGM_best}.
  }
\label{fig:GM_uninclined}
\end{figure}

\begin{figure}
\center
\includegraphics[scale=0.35,angle=-90]{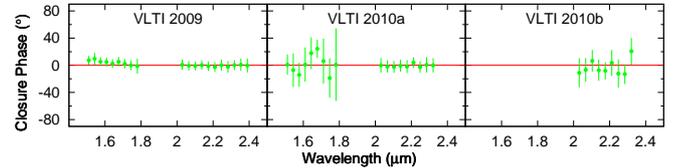}
\caption{
  Observed HD144432 closure phases as function of wavelength.
  }
\label{fig:obs_TGM_CP}
\end{figure}

\begin{table*}
\caption{The best-fit parameters of our geometric modeling.
$R_{\mathrm{\highlight{ring}}}$ is the inner radius of the ring-shaped disk models.
$f_{\mathrm{\highlight{ring}}}$, $f_{\mathrm{star}}$, and $f_{\mathrm{halo}}$
denote the flux contributions of the \highlight{ring}, the star, and the halo, respectively.
$i$ is the inclination angle ($0\degree$ corresponds to a face-on \highlight{ring}).
$\theta_{\mathrm D}$ is the position angle of the \highlight{ring}'s major axis.
$\chi^2_{\mathrm{red}}$ is the reduced chi-square.
}
\label{tab:GM}
\center
\begin{tabular}{ccccccccc}
\hline
\hline
Model  & Band 
              &$R_{\mathrm{\highlight{ring}}} (\AU)$
                                &$f_{\mathrm{\highlight{ring}}}$
                                        &$f_{\mathrm{star}}$
                                                              &$f_{\mathrm{halo}}$
                                                                                &$i (\degree)$
                                                                                              &$\theta_{\mathrm D} (\degree)$
                                                                                                              &$\chi^2_{\mathrm{red}}$ \\ 
\hline
star-\highlight{ring}
       & $K$    &$0.210\pm0.008$  &0.73   & 0.27                &0                &             &               &$ 2.3 $  \\
       & $H$    &$0.204\pm0.011$  &0.54   & 0.46                &0                &             &               &$ 3.4 $ \\
\hline
star-\highlight{ring}-halo
       & $K$    &$0.169\pm0.009$  &0.73   & $0.15 \pm0.02 $     &$0.12 \pm0.02 $  &             &               &$ 0.8 $    \\
       & $H$    &$0.180\pm0.010$  &0.54   & $0.40 \pm0.02 $     &$0.06 \pm0.02 $  &             &               &$ 1.6 $ \\
\hline
star-\highlight{ring}
       & $K$    &$0.223\pm0.015$  &0.73   & 0.27                &0                &$26  \pm12 $ &$-81  \pm16  $ &$ 2.5 $\\
(inclined)
       & $H$    &$0.218\pm0.018$  &0.54   & 0.46                &0                &$34  \pm16 $ &$17  \pm25  $  &$ 3.9 $\\
\hline
star-\highlight{ring}-halo
       & $K$    &$0.178\pm0.014$  &0.73   & $0.16 \pm0.03 $     &$0.11 \pm0.03 $  &$24  \pm10 $ &$-86  \pm16  $ &$ 0.8 $   \\
(inclined)
       & $H$    &$0.192\pm0.012$  &0.54   & $0.40 \pm0.02 $     &$0.06 \pm0.02 $  &$37  \pm11 $ &$ 1  \pm15 $   &$ 1.5 $\\
\hline
\end{tabular}
\end{table*}

\begin{table*}
\caption{The best-fit parameters of the {\TGM}s.} 
\label{tab:TGM_best}
\setlength{\tabcolsep}{4pt}
\renewcommand{\arraystretch}{1.4}
\center
\begin{tabular}{cccccccccccccccc}
\hline
\hline
&$k_\mathrm{halo}$          &   $i(\degree)$          &  $\theta_{D}(\degree)$
& $T_{\mathrm{1,in}}($K$)$&   $q_1$                   &$r_{\mathrm{1,in}}(\AU)$ & $\Delta r_1(\AU)$
& $T_{\mathrm{2,in}}($K$)$&   $q_2$                   &$r_{\mathrm{2,in}}(\AU)$ & $\Delta r_2(\AU)$
& $\chi^2_\mathrm{red}$
\\
\hline
one-component
& $0.306             $      &  $60          $               &  $33                  $
& $800              $   &  $0.63$                   & $0.392                  $     & $5.0                    $
&                       &                           &                               &                         
& 10.7
\\
\hline
two-component
& $0.185_{-0.013 }^{+0.013}$&  $10_{-10}^{+8}$ \tablefootmark{a}          &  $30  _{-120 }^{+60  }$
& $1567_{-29 }^{+55}$   &  $0.5$\tablefootmark{b}   & $0.214_{-0.003}^{+0.005}$     & $0.019_{-0.002}^{+0.001}$
& $ 392_{-7  }^{+6 }$   &  $0.85_{-0.04}^{0.01 }$   & $0.93 _{-0.03 }^{+0.02 }$     & $7.9  _{-0.5  }^{+0.5  }$
           & 1.15
\\
\hline
\end{tabular}
\tablefoot{
  \\ \tablefoottext{a}{With 99.7\% confident limit $i<28\degree$.}
  \\ \tablefoottext{b}{
The power law index $q_1$ was fixed to $0.5$ 
\citep[corresponding to a flared irradiated disk;][]{Kenyon1987},
since $q_1$ cannot be constrained because of the small radial thickness of the inner disk ring (see Appendix A.3).
 }
}
\end{table*}

In this section, we attempt to build models that can reproduce both the data from our new observations and data sets available in literature,
including the $K$-band observation with KI \citep[M05,][]{Eisner2009}, the $H$-band observation with IOTA (M06),
the MIR interferometry with VLTI/MIDI \citep{Leinert2004},
as well as the SED \citep{Leinert2004}.
In the modeling process, we assume a distance of $145~\mathrm{pc}$ \citep{Perez2004}.

\subsection{Geometric modeling}

First, simple geometric models are employed to characterize the size of the disk.
In order to be consistent with M05's
work, we used the uniform-brightness ring model, which has a ring
thickness of 20\% of the inner radius $R_{\mathrm{\highlight{ring}}}$.
By fitting a Kurucz model
\citep{Kurucz1992} to the stellar component of the dereddened SED (taken
from Leinert et al. 2004) and measuring the flux excess in the $H$ and $K$ bands,
we obtained the \highlight{ring (disk)} flux fractions $f_{\mathrm{\highlight{ring}}}=0.54$ for the $H$
band and $f_{\mathrm{\highlight{ring}}}=0.73$ for the $K$ band, respectively.

The wavelength-averaged $H$- and $K$-band visibilities were fitted with a ring model including
the stellar contribution.
The total visibility of the star-\highlight{ring model} can be described by:
\begin{equation}
V_{\mathrm{total}} = f_{\mathrm{\highlight{ring}}} V_{\mathrm{\highlight{ring}}} + f_{*}V_{*},
\label{visibility}
\end{equation}
where $V_{\mathrm{total}}$ is the measured visibility,
$f_{*}=1-f_{\mathrm{\highlight{ring}}}$ is the flux contribution from the central star,
and $V_{*}$ is the stellar visibility. 
As the central star is unresolved ($R_*<0.1~\mas$), we set $V_{*}=1$.
\highlight{
The visibility $V_\mathrm{ring}$ of an uniform ring is calculated following Eq. (8) in \citet{Eisner2004}.
}
The best-fit geometric star-\highlight{ring} models are presented in Fig. 1 and Table \ref{tab:GM} \citep[for a distance of $145~\pc$,][]{Perez2004}.
We derived ring-fit radii of $R_{\mathrm{\highlight{ring}}}= 0.21 \pm0.01 ~\AU$ for the $K$ band
and $R_{\mathrm{\highlight{ring}}}=0.20\pm0.01~\AU$ for the $H$ band.

The high reduced chi-square errors $\chi^2_{\mathrm{red}}$ of the visibility fits
(see last column in Table \ref{tab:GM}) indicate that even the best-fit
star-\highlight{ring} models
cannot reproduce the observational data well.
M06 reported that for some Herbig Ae stars an additional overresolved halo component
(i.e., a halo that is too large to be constrained by the short baselines)
is required for fitting the visibilities with simple geometric models.
Therefore, we introduce a halo component into our model.
We assume that the halo structure scatters the light from the central star.
While $f_{\mathrm{\highlight{ring}}}$ remains the same parameter as above,
$f_{*}$ is given by $f_{*}=1-f_{\mathrm{\highlight{ring}}}-f_{\mathrm{halo}}$,
where the halo flux ratio, $f_{\mathrm{halo}}$, is an additional free parameter.
The best-fit star-\highlight{ring}-halo model in the $K$ band shows that $12\pm2~\%$ of the flux is emitted by the halo,
and that the \highlight{ring} has a ring-fit radius of $R_{\mathrm{\highlight{ring}}}=0.17\pm0.01~\AU$.
In the $H$ band, the halo contributes $ 6\pm2~\%$ to the total flux and the
ring-fit radius is $0.18\pm0.01~\AU$.
With an additional halo component, the values of the $\chi^2_{\mathrm{red}}$ decrease by a factor of ${\sim}2.5$ (see Table \ref{tab:GM}).

The measurements at different position angles allow us to investigate the
inclination of the disk.
Thus, we also employed inclined ring models with and without an extended halo.
The fitting results are summarized in Table \ref{tab:GM}.
The best-fit models suggest small inclination angles of ${\sim}30^{\circ}$,
but the $\chi^2_{\mathrm{red}}$ values do not show a significant decrease.

\highlight{
Finally, to roughly characterize the N-band size,
we fitted the band-averaged visibility of $0.19\pm0.1$
with a thin-ring model and derived a ring radius of $\sim1.4~\AU$.
We also studied a geometric two-ring model and derived a fit radius of ~1.6 AU for the outer ring,
when we assumed an estimated N-band flux contribution of $15\%$ from the inner model ring
and an inner-ring radius of 0.18 AU, as derived in the above star-ring-halo model (see Table 2).
}

\begin{figure}
\flushleft
\center
\raisebox{-1pc}{\makebox[0cm][r]{a}}%
\mbox{\includegraphics[scale=0.35,angle=-90]{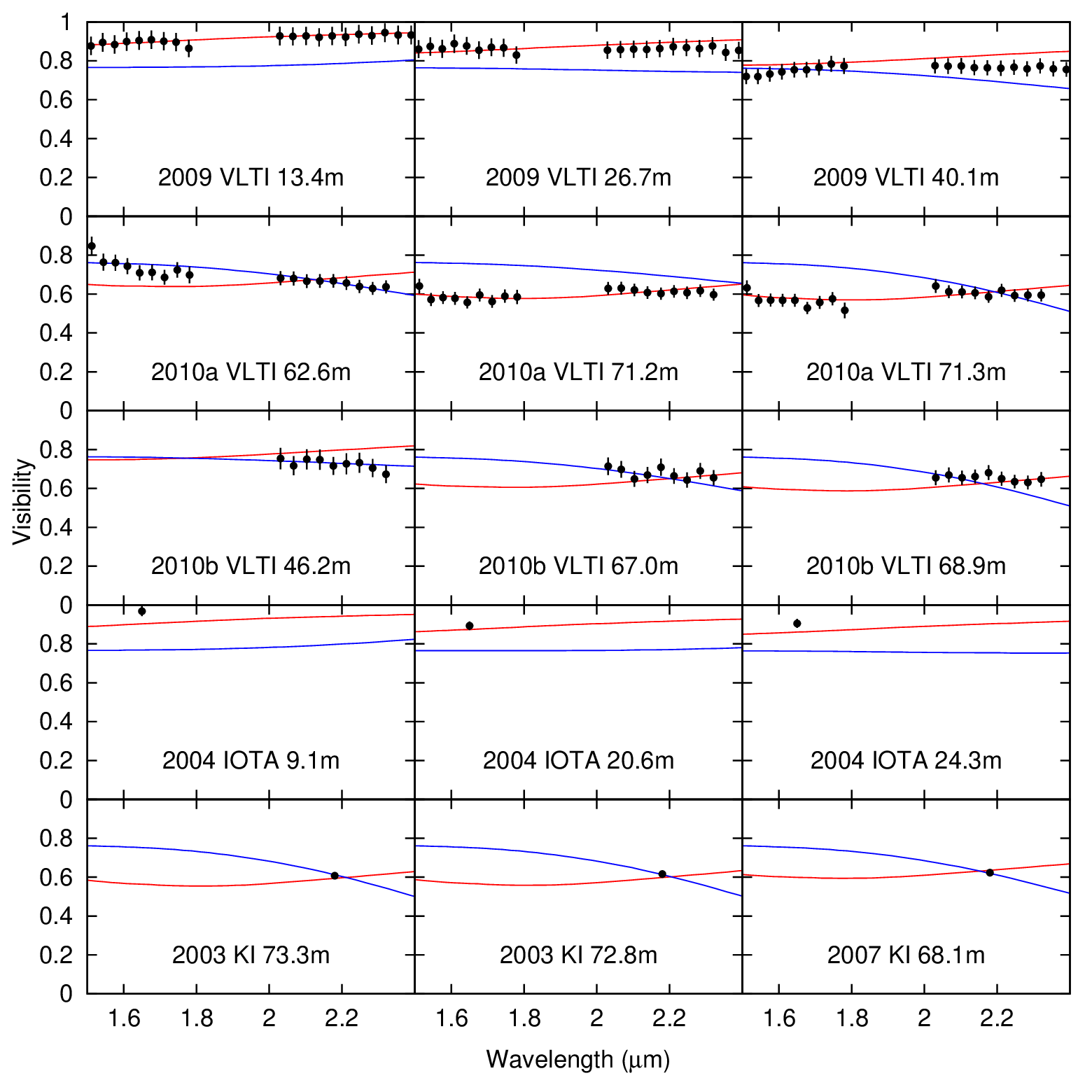}}
\vbox{\includegraphics[scale=0.35,angle=-90]{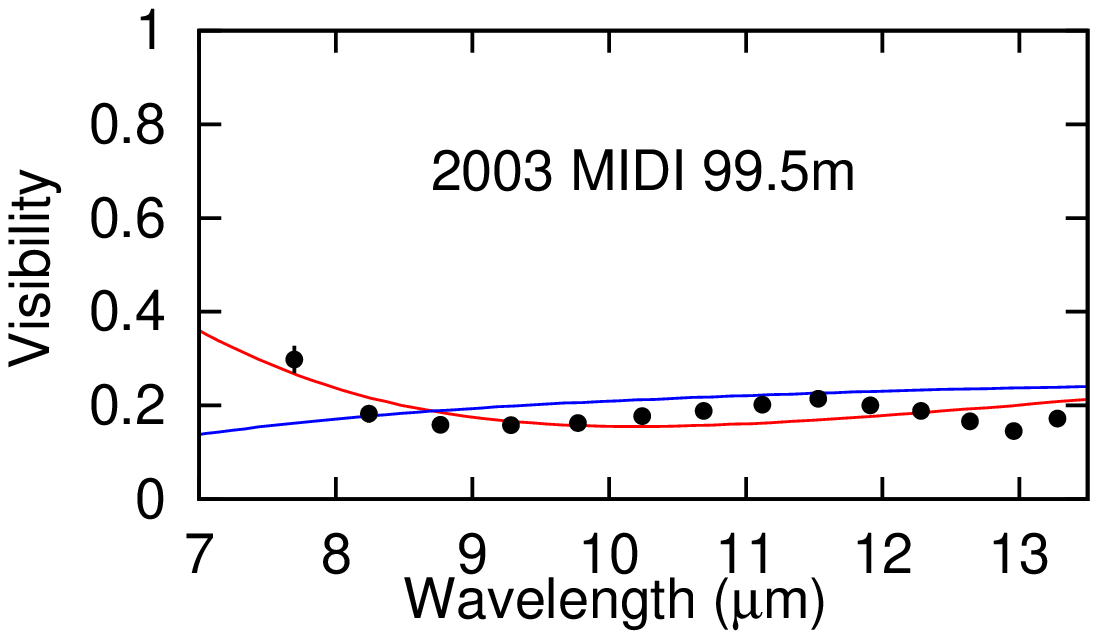}}
\raisebox{-1pc}{\makebox[0cm][r]{b}}%
\mbox{\includegraphics[scale=0.35,angle=-90]{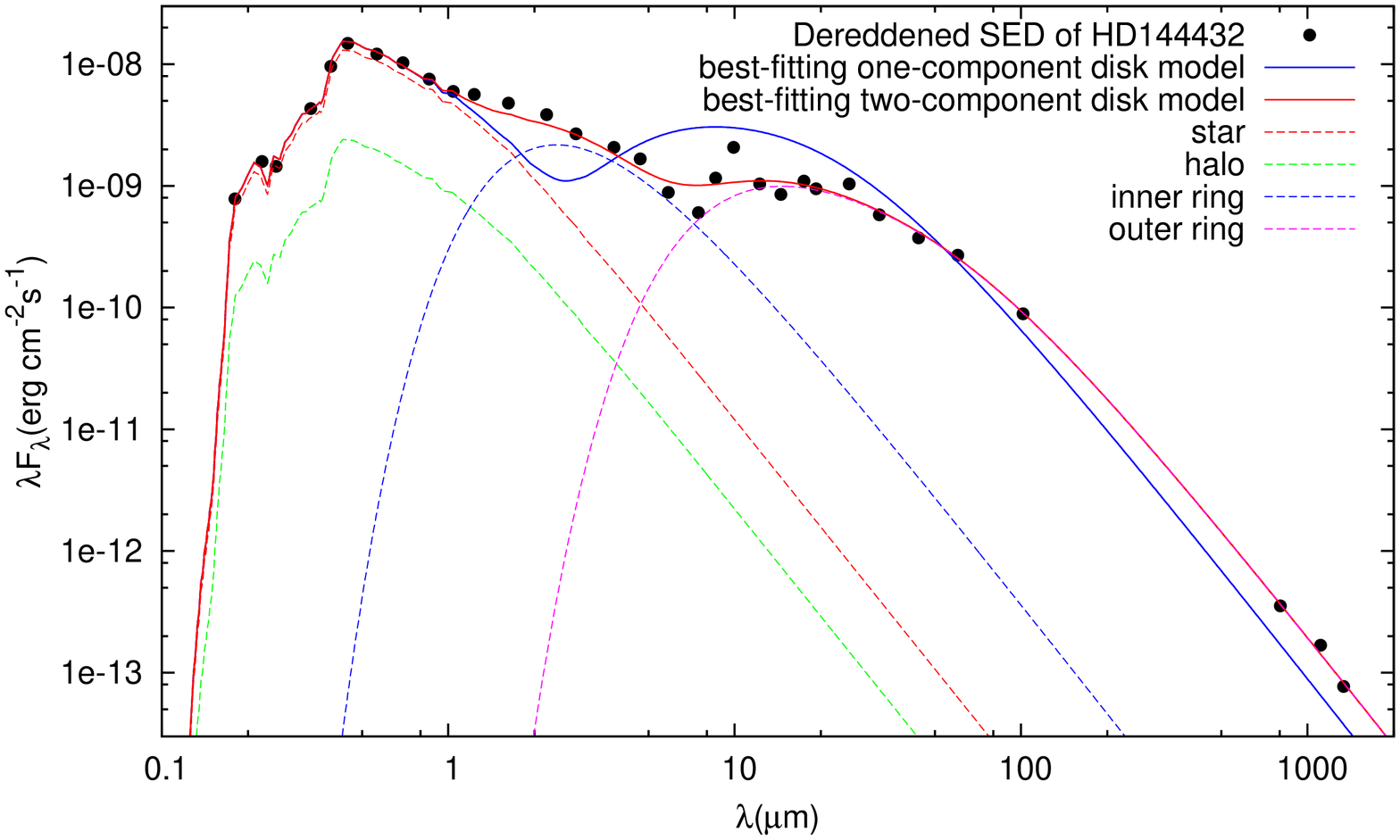}}
\raisebox{-1pc}{\makebox[0cm][r]{c}}%
\mbox{\includegraphics[scale=0.35,angle=-90]{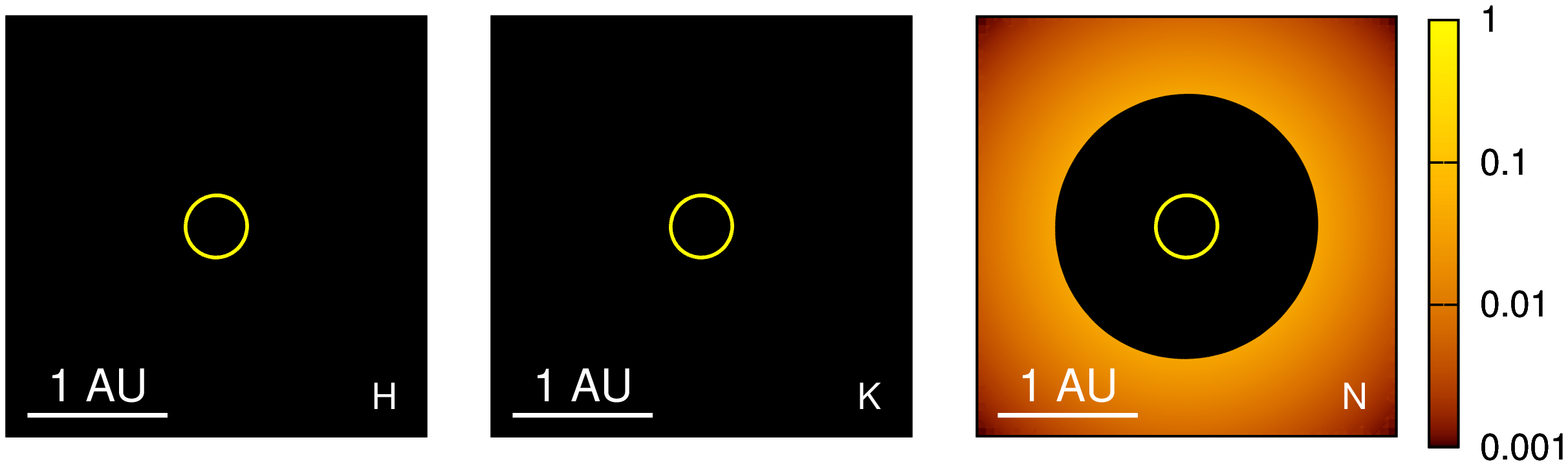}}

\iftrue
\caption{
  Comparison of the best-fit two-component disk model (red lines; see Table \ref{tab:TGM_best})
  and best-fit one-component disk model (blue lines)
  with the observations of HD144432.
  {\it a:} near- and mid-infrared visibilities.
  The data sets 2004 IOTA, 2003 KI, 2007 KI, and 2003 MIDI mentioned in the panels
  are taken from
  \citet{Monnier2006}, \citet{Monnier2005}, \citet{Eisner2009}, and \citet{Leinert2004},
  respectively.
  {\it b:} SED \citep[from][]{Leinert2004} and {\TGM}s.
  The dashed lines denote the contributions from individual components in the best-fit two-component disk model.
  {\it c:} 
  $H$-, $K$-, and $N$-band intensity distributions of the best-fit two-component disk model (Table 3).
  The second component of the disk, i.e., the more extended part, is visible only in the $N$ band.
  The star and halo are not plotted.
  }
\fi

\label{fig:obs_TGM}
\label{fig:TGM}
\label{fig:image}
\end{figure}

\subsection{Temperature-gradient modeling with a one-component disk}
Temperature-gradient models are employed to fit the NIR and MIR visibilities as well as the SED data simultaneously.
We begin with a simple model including a star, an inclined one-component disk and an extended overresolved halo.
The star contributes with the flux $F_*(\lambda)=A_* I_*(\lambda)$,
where  $I_*(\lambda)$ is the intensity at the stellar surface,
and $A_*=\pi R_*^2$ is the angular area of the star.
The halo is assumed to be very extended, emitting with a spectrum similar to the stellar spectrum.
Thus its flux is  $F_{\mathrm{halo}}(\lambda)=k_{\mathrm{halo}}F_*(\lambda)$.
The disk is an optically thick ring with a power-law temperature distribution \citep{Hillenbrand1992}
\begin{math}
T= T_{\mathrm{in}}
\left(
{r}/{r_\mathrm{in}}
\right)^{-q}
,~r_{\mathrm{in}}\leq r \leq r_{\mathrm{out}}
\end{math},
with the inner and outer radius $r_{\mathrm{in}}$ and $r_{\mathrm{out}}$ of the disk,
a disk temperature $T_{\mathrm{in}}$ at $r_{\mathrm{in}}$,
and a power-law index $q$.
Each part of the ring emits black body radiation at its local temperature.
We first fitted the SED
\highlight{in the UV/optical band ($0.1\enDash1~\mu\mathrm{m}$) }
with a Kurucz model \citep{Kurucz1992}
in order to determine the stellar parameters to be used in our further modeling,
i.e., temperature $T_*=7180~\mathrm{K}$, surface gravity $\log g=2.78$, and metallicity $\log m=-0.415$.
The seven free parameters in the model are the inclination angle $i$,
the position angle $\theta_{\mathrm D}$ of the disk's major axis, $T_{\mathrm{in}}$,  $q$,
$r_{\mathrm{in}}$, the radial thickness $\Delta r=r_{\mathrm{out}}-r_{\mathrm{in}}$ of the ring, and the flux fraction $k_{\mathrm{halo}}$.

We tried to find the best-fit model by scanning the physically reasonable parameter range
(calculation of approximately $6\times10^7$ models;
details of the searching process is described in the Appendix A).
Unfortunately, we did not find a model that can  reproduce the data reasonably well.
The best-fit model (see blue model lines in Fig. \ref{fig:obs_TGM})
deviates much from the data,
and its $\chi^2_{\mathrm{red}}$ is $10.7$.
This deviation suggests a more sophisticated structure of the disk.

\subsection{Temperature-gradient modeling with a two-component disk}
Since the $\chi^2_{\mathrm{red}}$ of the one-component disk model discussed above is very large,
we introduced a slightly more complicated model in a second modeling step,
in which the disk consists of two power-law components,
with temperature
$ T= T_{\mathrm{1,in}} \left( {r}/{r_\mathrm{1,in}} \right)^{-q_1} $
for
$~r_{\mathrm{1,in}}\leq r \leq r_{\mathrm{1,out}} $
and
$ T= T_{\mathrm{2,in}} \left( {r}/{r_\mathrm{2,in}} \right)^{-q_2} $
for
$~r_{\mathrm{2,in}}\leq r \leq r_{\mathrm{2,out}} $%
, respectively.
We use the subscript 1 to denote the inner disk,
and 2 for the outer disk.
The assumptions for the star and halo remain unchanged.
After computation of several $10^{10}$ models,
we found the best-fit solution shown in Table \ref{tab:TGM_best} and Fig. \ref{fig:obs_TGM}
(with $\chi^2=1.15$; see Appendix A for the searching process).

This best-fit model consists of the star, a halo with a brightness of $18\%$ of the star,
and a nearly face-on ($i<28\degree$) two-component disk.
The inner part of the disk is a thin ring at $r_{\mathrm{1,in}}\sim0.21~\AU$,
with a temperature $T_\mathrm{1,in}\sim1600~\mathrm{K}$, and a radial thickness $\Delta r_1 \sim 0.02~\AU$.
The power law index $q_1$ was fixed to $0.5$ 
\citep[corresponding to a flared irradiated disk;][]{Kenyon1987},
since $q_1$ cannot be constrained because of the small radial thickness of the inner disk ring (see Appendix A.3).
The outer part extends from ${\sim}1~\AU$ to ${\sim}10~\AU$, with an inner temperature ${\sim}400~\mathrm{K}$.
The contribution from the components to the total NIR flux (at $2~\mu\mathrm{m}$)
are $60\%$ from the inner disk, $33\%$ from the central star, and $6\%$ from the halo, and almost zero from the outer disk.
The contribution from the components to the total MIR flux (at $10~\mu\mathrm{m}$)
are $21\%$ from the inner disk, $78\%$ from the outer disk, and only $1\%$ from the star and halo.
In Fig. \ref{fig:obs_TGM}
we show the comparison of this best-fit model ($\chi^2_\mathrm{red}=1.15$)
to our AMBER visibilities
together with visibilites from IOTA, Keck, and MIDI measurements,
as well as SED data from literature.

\highlight{
A prominent feature of the model is the large gap between the inner and outer disk.
The size of the NIR emitting region is confined to $\sim0.21\pm0.01~\AU$ by the NIR visibilities.
The low visibilities in the MIR suggest a much larger size of $\gtrsim1~\AU$ for the MIR emitting region.
If a smooth temperature profile (and hence a continuous emission distribution)
is assumed between the two distinct length scales,
the NIR emitting region will be much broader than in our model,
and would lead to a NIR flux much higher than the observation.
Therefore, the gap in our model is strongly required for intepreting all data simultaneously.
}

\section{Discussion}
\begin{figure}
\includegraphics[scale=0.35,angle=-90]{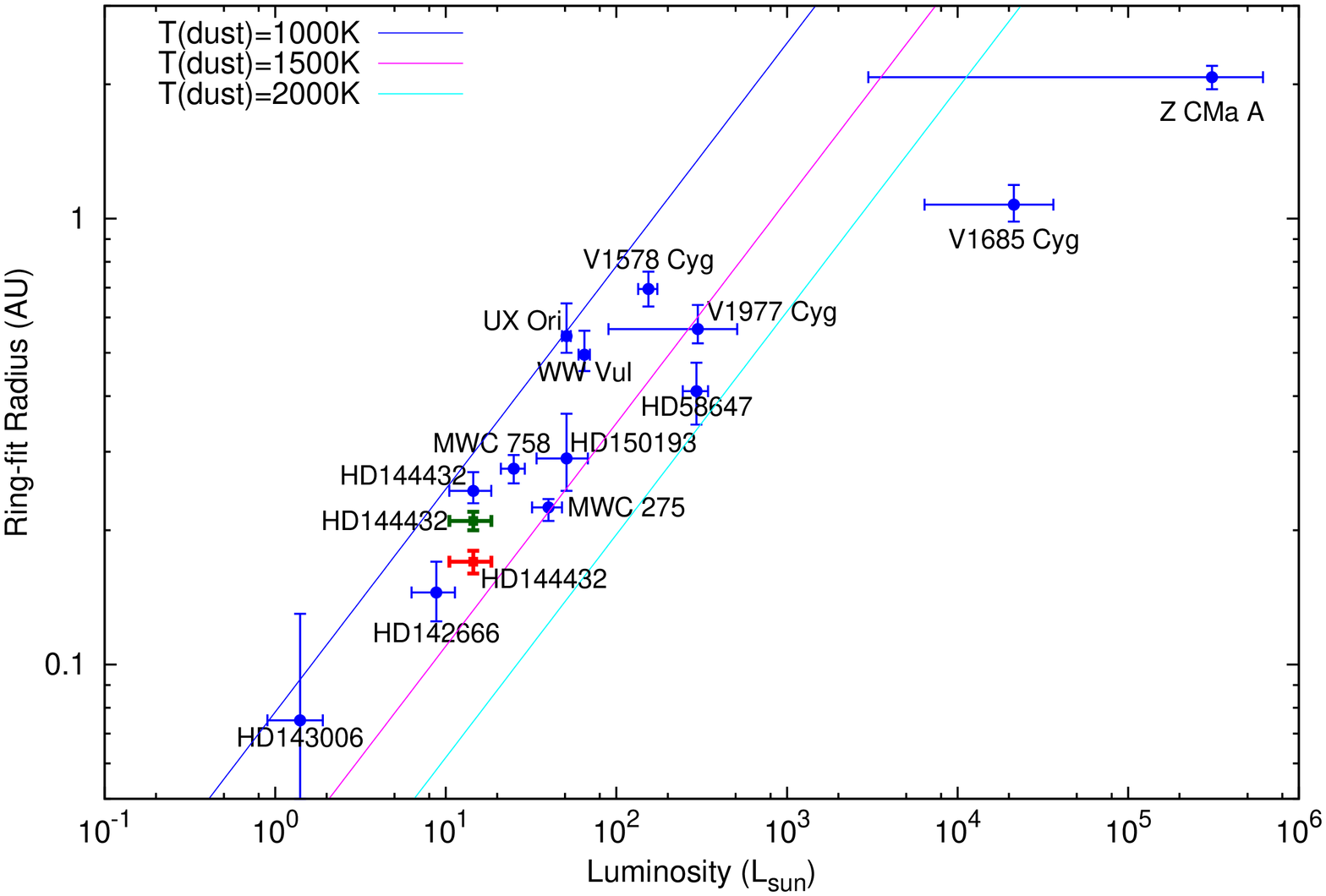}
\caption{
  Size-Luminosity diagram for HAeBe stars \citep{Monnier2002}.
  Blue dots: data taken from \citet{Monnier2005}.
  Red square: HD144432 (ring-fit $K$-band radius of the star-ring-halo uninclined model).
  Dark-green square: HD144432 (ring-fit $K$-band radius of the star-ring uninclined model).
  The error bars of the radii for HD144432 are only the uncertainty from visibilty measurements,
  and do not include the distance uncertainty.
  Lines: Theoretical dust sublimation radius for gray dust and three different dust sublimation temperatures.
  Predictions of models including backwarming are discussed in Sect.4.
  }
\label{fig:size_lumi}
\end{figure}

To compare the derived NIR sizes with the expected dust sublimation radius
and with other HAEBE stars,
we plot HD144432 into the size-luminosity diagram introduced by \citet{Monnier2002}
using our $K$-band ring-fit radii
(inclination $i=0\degree$ models, see Sect. 3.1 and Table 1)
together with the luminosity $L=14.5\pm4~L_\odot$ (M05).
Figure \ref{fig:size_lumi} shows that the inner ring radius of
$0.17\pm0.01~\AU$ (star-disk-halo model; adopting a distance of ${\sim}145\pm20~\pc$, see below)
is roughly consistent with the predicted dust-sublimation radius of $0.13~\AU$
corresponding to a sublimation temperature of $1500~\mathrm{K}$
and gray dust opacities.
Furthermore, we compared the measured radius with the prediction
of a model including backwarming and accretion luminosity \citep{Millan-Gabet2007}.
This model suggests an inner disk radius of ${\sim}0.27~\AU$
for a stellar luminosity of $14.5~L_\odot$ and an accretion luminosity of $1.0~L_\odot$
(\citealp{GarciaLopez2006}; we scaled down the value according to the difference in assumed star parameters).
Our measured inner ring radius of ${\sim}0.17~\AU$ lies in the range between these two model predictions
of $0.13~\AU$ and $0.27~\AU$.

The error bars of the radii for HD144432 in Fig. \ref{fig:size_lumi} are only the uncertainty from visibility measurements
and do not include the uncertainty of the distance.
\citet{Perez2004} conclude that HD144432 is likely a member of the star association Sco OB 2-2, with a distance of ${\sim}145~\pc$. 
However, in previous studies, distances in range from $108~\pc$ to $2.4~{\rm kpc}$ were also reported \citep[e.g.,][]{Pottasch1988,Perez2004}.
In this paper, we adopted the distance of $145\pm20~\pc$
\citep[the error bar corresponds to the distance dispersion within Sco OB 2-2; see][]{Preibisch2002}.
In spite of the distance uncertainty, it is possible to discuss the location of HD144432
relative to the $1500~\mathrm{K}$ line for the following reason.
If the adopted distance is wrong by a certain factor,
then this factor changes both the ring-fit radius and the luminosity
in such a way that the location of the star moves parallel to the sublimation radius lines
in the size-lumlinosity relation \citep{Monnier2002}.

Both our geometric and {\TGM}ing suggest the existence of a halo component,
which contributes ${\sim}6\enDash12\%$ to the total NIR flux in the best-fit {\GM}
(in the best-fit {\TGM}: ${\sim}6\%$ at $2~\mu\mathrm{m}$).
The size of the halo component cannot be measured precisely,
but the visibilities suggest that it is ${\gtrsim}1~\AU$.
Given its large distance to the central star,
the halo emission is probably dominated by scattered stellar light \citep[M06]{Akeson2005}.
Plausible origins of the halo material include an infalling remnant envelope
or dust entrained in the stellar wind/outflow (M06),
or the flaring outer disk,
which scatters the stellar light \citep{Pinte2008}.

Our {\TGM}ing suggests that
the disk consists of two components with distinct length scales.
The inner component is a thin ring at an inner radius of ${\sim}0.21~\AU$
with a temperature of ${\sim}1600~\mathrm{K}$
and a radial thickness ${\sim}0.02~\AU$.
The outer part extends from ${\sim}1~\AU$ to ${\sim}10~\AU$
with an inner temperature of ${\sim}400~\mathrm{K}$.
The disk is seen roughly face-on with an inclination angle of $i<28\degree$.

The small radial thickness of the inner ring-shaped disk
is consistent with the puffed-up rim model
\citep{Natta2001,Dullemond2001,Dullemond2002,Dominik2003,Dullemond2004a}.
In such a model, the NIR emission is dominated by the puffed-up inner rim at the dust sublimation radius,
and the region of the disk behind the rim will be colder due to shadowing effects,
which can lead to a gap in the observed intensity distribution.
A gap phenomenon was also reported for other Herbig stars
\citep[e.g.,][]{Benisty2010,Tatulli2011}.
While the NIR emission of the disk is dominated exclusively by the inner ring (presumably the puffed-up rim),
the MIR emission has a bimodal distribution, with ${\sim}20\%$ from the inner ring,
and the rest from the outer part.
Such a spatial distribution of MIR emission is also consistent with the prediction of the rim model \citep{vanBoekel2005}.

\section{Summary and conclusions}
We presented VLTI/AMBER observations of the Herbig Ae star HD144432 in the $H$ and $K$ bands.
The following results were obtained.

The $K$- and $H$-band emitting regions have geometric ring-fit radii of
$0.21\pm0.01~\AU$ and $0.20\pm0.01~\AU$, respectively.
If we introduce an additional halo component,
we obtain the smaller disk ring-fit radii of
$0.17\pm0.01~\AU$ ($K$ band) and $0.18\pm0.01~\AU$ ($H$ band).
This measured $K$-band ring radius of ${\sim}0.17~\AU$ lies in the range between
the above discussed dust sublimation radius of ${\sim}0.13~\AU$
(size-luminosity relation in Fig.4; for sublimation temperature of $1500~\mathrm{K}$ and gray dust),
and the prediction of models including backwarming (${\sim}0.27~\AU$).
Both our geometric and {\TGM}ing indicate the existence of an additional extended halo component.
In the best-fit {\TGM},
the halo contributes ${\sim}6\%$ at $2~\mu\mathrm{m}$.

Our best-fit temperature-gradient disk model can approximately reproduce both the NIR-MIR visibilities and the SED data (Fig. 3).
The model consists of the central star, an extended halo, and a nearly face-on two-component disk.
In the modeling procedure, we started with a very wide range for all model parameters,
calculated several $10^{10}$ models corresponding to all combination of the parameters,
and finally obtained disk parameters that seem to be physically quite reasonable.
The inner part of the disk is a thin ring at an inner radius of ${\sim}0.21~\AU$
with a temperature of ${\sim}1600~\mathrm{K}$
and a radial thickness ${\sim}0.02~\AU$.
The outer part extends from ${\sim}1~\AU$ to ${\sim}10~\AU$
with an inner temperature of ${\sim}400~\mathrm{K}$.
The NIR emission of the disk is dominated by the inner ring.
The MIR emission has a bimodal distribution, with ${\sim}20\%$ from the inner ring
and the rest from the outer part.
The {\TGM}ing suggests an upper limit for the inclination angle of $i<28\degree$.

\begin{acknowledgements}
\highlight{
We thank the ESO VLTI team on Paranal for the excellent collaboration.
The data presented here were reduced using the publicly available data-reduction software package amdlib
kindly provided by the Jean-Marie Mariotti Center (http://www.jmmc.fr/datar\_processing\_amber.htm).
This publication makes use of the SIMBAD database operated at CDS, Strasbourg, France.
Finally, we thank the anonymous referee for his helpful comments.
}
\end{acknowledgements}

\bibliography{bib/HD144432,bib/YSOdiskTheory,bib/other}

\begin{thebibliography}{37}
\expandafter\ifx\csname natexlab\endcsname\relax\def\natexlab#1{#1}\fi

\bibitem[{{Acke} \& {van den Ancker}(2004)}]{Acke2004}
{Acke}, B. \& {van den Ancker}, M.~E. 2004, \aap, 426, 151

\bibitem[{{Akeson} {et~al.}(2005){Akeson}, {Walker}, {Wood}, {Eisner}, {Scire},
  {Penprase}, {Ciardi}, {van Belle}, {Whitney}, \& {Bjorkman}}]{Akeson2005}
{Akeson}, R.~L., {Walker}, C.~H., {Wood}, K., {et~al.} 2005, \apj, 622, 440

\bibitem[{{Benisty} {et~al.}(2010){Benisty}, {Tatulli}, {M{\'e}nard}, \&
  {Swain}}]{Benisty2010}
{Benisty}, M., {Tatulli}, E., {M{\'e}nard}, F., \& {Swain}, M.~R. 2010, \aap,
  511, A75+

\bibitem[{{Carmona} {et~al.}(2007){Carmona}, {van den Ancker}, \&
  {Henning}}]{Carmona2007}
{Carmona}, A., {van den Ancker}, M.~E., \& {Henning}, T. 2007, \aap, 464, 687

\bibitem[{{Dominik} {et~al.}(2003){Dominik}, {Dullemond}, {Waters}, \&
  {Walch}}]{Dominik2003}
{Dominik}, C., {Dullemond}, C.~P., {Waters}, L.~B.~F.~M., \& {Walch}, S. 2003,
  \aap, 398, 607

\bibitem[{{Dullemond}(2002)}]{Dullemond2002}
{Dullemond}, C.~P. 2002, \aap, 395, 853

\bibitem[{{Dullemond} \& {Dominik}(2004)}]{Dullemond2004a}
{Dullemond}, C.~P. \& {Dominik}, C. 2004, \aap, 417, 159

\bibitem[{{Dullemond} {et~al.}(2001){Dullemond}, {Dominik}, \&
  {Natta}}]{Dullemond2001}
{Dullemond}, C.~P., {Dominik}, C., \& {Natta}, A. 2001, \apj, 560, 957

\bibitem[{{Eisner} {et~al.}(2009){Eisner}, {Graham}, {Akeson}, \&
  {Najita}}]{Eisner2009}
{Eisner}, J.~A., {Graham}, J.~R., {Akeson}, R.~L., \& {Najita}, J. 2009, \apj,
  692, 309

\bibitem[{{Eisner} {et~al.}(2004){Eisner}, {Lane}, {Hillenbrand}, {Akeson}, \&
  {Sargent}}]{Eisner2004}
{Eisner}, J.~A., {Lane}, B.~F., {Hillenbrand}, L.~A., {Akeson}, R.~L., \&
  {Sargent}, A.~I. 2004, \apj, 613, 1049

\bibitem[{{Garcia Lopez} {et~al.}(2006){Garcia Lopez}, {Natta}, {Testi}, \&
  {Habart}}]{GarciaLopez2006}
{Garcia Lopez}, R., {Natta}, A., {Testi}, L., \& {Habart}, E. 2006, \aap, 459,
  837

\bibitem[{{Hillenbrand} {et~al.}(1992){Hillenbrand}, {Strom}, {Vrba}, \&
  {Keene}}]{Hillenbrand1992}
{Hillenbrand}, L.~A., {Strom}, S.~E., {Vrba}, F.~J., \& {Keene}, J. 1992, \apj,
  397, 613

\bibitem[{{Keller} {et~al.}(2008){Keller}, {Sloan}, {Forrest}, {Ayala},
  {D'Alessio}, {Shah}, {Calvet}, {Najita}, {Li}, {Hartmann}, {Sargent},
  {Watson}, \& {Chen}}]{Keller2008}
{Keller}, L.~D., {Sloan}, G.~C., {Forrest}, W.~J., {et~al.} 2008, \apj, 684,
  411

\bibitem[{{Kenyon} \& {Hartmann}(1987)}]{Kenyon1987}
{Kenyon}, S.~J. \& {Hartmann}, L. 1987, \apj, 323, 714

\bibitem[{{Kreplin, A.} {et~al.}(2012){Kreplin, A.}, {Kraus, S.}, {Hofmann,
  K.-H.}, {Schertl, D.}, {Weigelt, G.}, \& {Driebe, T.}}]{Kreplin2011}
{Kreplin, A.}, {Kraus, S.}, {Hofmann, K.-H.}, {et~al.} 2012, A\&A, 537, A103

\bibitem[{{Kurucz}(1992)}]{Kurucz1992}
{Kurucz}, R.~L. 1992, in IAU Symposium, Vol. 149, The Stellar Populations of
  Galaxies, ed. {B.~Barbuy \& A.~Renzini}, 225--+

\bibitem[{{Leinert} {et~al.}(2004){Leinert}, {van Boekel}, {Waters},
  {Chesneau}, {Malbet}, {K{\"o}hler}, {Jaffe}, {Ratzka}, {Dutrey}, {Preibisch},
  {Graser}, {Bakker}, {Chagnon}, {Cotton}, {Dominik}, {Dullemond},
  {Glazenborg-Kluttig}, {Glindemann}, {Henning}, {Hofmann}, {de Jong},
  {Lenzen}, {Ligori}, {Lopez}, {Meisner}, {Morel}, {Paresce}, {Pel},
  {Percheron}, {Perrin}, {Przygodda}, {Richichi}, {Sch{\"o}ller}, {Schuller},
  {Stecklum}, {van den Ancker}, {von der L{\"u}he}, \& {Weigelt}}]{Leinert2004}
{Leinert}, C., {van Boekel}, R., {Waters}, L.~B.~F.~M., {et~al.} 2004, \aap,
  423, 537

\bibitem[{{Malfait} {et~al.}(1998){Malfait}, {Bogaert}, \&
  {Waelkens}}]{Malfait1998}
{Malfait}, K., {Bogaert}, E., \& {Waelkens}, C. 1998, \aap, 331, 211

\bibitem[{{Meeus} {et~al.}(1998){Meeus}, {Waelkens}, \& {Malfait}}]{Meeus1998}
{Meeus}, G., {Waelkens}, C., \& {Malfait}, K. 1998, \aap, 329, 131

\bibitem[{{Meeus} {et~al.}(2001){Meeus}, {Waters}, {Bouwman}, {van den Ancker},
  {Waelkens}, \& {Malfait}}]{Meeus2001}
{Meeus}, G., {Waters}, L.~B.~F.~M., {Bouwman}, J., {et~al.} 2001, \aap, 365,
  476

\bibitem[{{Millan-Gabet} {et~al.}(2007){Millan-Gabet}, {Malbet}, {Akeson},
  {Leinert}, {Monnier}, \& {Waters}}]{Millan-Gabet2007}
{Millan-Gabet}, R., {Malbet}, F., {Akeson}, R., {et~al.} 2007, Protostars and
  Planets V, 539

\bibitem[{{Monnier} {et~al.}(2006){Monnier}, {Berger}, {Millan-Gabet}, {Traub},
  {Schloerb}, {Pedretti}, {Benisty}, {Carleton}, {Haguenauer}, {Kern},
  {Labeye}, {Lacasse}, {Malbet}, {Perraut}, {Pearlman}, \&
  {Zhao}}]{Monnier2006}
{Monnier}, J.~D., {Berger}, J., {Millan-Gabet}, R., {et~al.} 2006, \apj, 647,
  444

\bibitem[{{Monnier} \& {Millan-Gabet}(2002)}]{Monnier2002}
{Monnier}, J.~D. \& {Millan-Gabet}, R. 2002, \apj, 579, 694

\bibitem[{{Monnier} {et~al.}(2005){Monnier}, {Millan-Gabet}, {Billmeier},
  {Akeson}, {Wallace}, {Berger}, {Calvet}, {D'Alessio}, {Danchi}, {Hartmann},
  {Hillenbrand}, {Kuchner}, {Rajagopal}, {Traub}, {Tuthill}, {Boden}, {Booth},
  {Colavita}, {Gathright}, {Hrynevych}, {Le Mignant}, {Ligon}, {Neyman},
  {Swain}, {Thompson}, {Vasisht}, {Wizinowich}, {Beichman}, {Beletic},
  {Creech-Eakman}, {Koresko}, {Sargent}, {Shao}, \& {van Belle}}]{Monnier2005}
{Monnier}, J.~D., {Millan-Gabet}, R., {Billmeier}, R., {et~al.} 2005, \apj,
  624, 832

\bibitem[{{Monnier} {et~al.}(2009){Monnier}, {Tuthill}, {Ireland}, {Cohen},
  {Tannirkulam}, \& {Perrin}}]{Monnier2009}
{Monnier}, J.~D., {Tuthill}, P.~G., {Ireland}, M., {et~al.} 2009, \apj, 700,
  491

\bibitem[{{Natta} {et~al.}(2001){Natta}, {Prusti}, {Neri}, {Wooden}, {Grinin},
  \& {Mannings}}]{Natta2001}
{Natta}, A., {Prusti}, T., {Neri}, R., {et~al.} 2001, \aap, 371, 186

\bibitem[{{Pasinetti Fracassini} {et~al.}(2001){Pasinetti Fracassini},
  {Pastori}, {Covino}, \& {Pozzi}}]{Pasinetti-Fracassini2001}
{Pasinetti Fracassini}, L.~E., {Pastori}, L., {Covino}, S., \& {Pozzi}, A.
  2001, \aap, 367, 521

\bibitem[{{P{\'e}rez} {et~al.}(2004){P{\'e}rez}, {van den Ancker}, {de Winter},
  \& {Bopp}}]{Perez2004}
{P{\'e}rez}, M.~R., {van den Ancker}, M.~E., {de Winter}, D., \& {Bopp}, B.~W.
  2004, \aap, 416, 647

\bibitem[{{Petrov} {et~al.}(2007){Petrov}, {Malbet}, {Weigelt}, {Antonelli},
  {Beckmann}, {Bresson}, {Chelli}, {Dugu{\'e}}, {Duvert}, {Gennari},
  {Gl{\"u}ck}, {Kern}, {Lagarde}, {Le Coarer}, {Lisi}, {Millour}, {Perraut},
  {Puget}, {Rantakyr{\"o}}, {Robbe-Dubois}, {Roussel}, {Salinari}, {Tatulli},
  {Zins}, {Accardo}, {Acke}, {Agabi}, {Altariba}, {Arezki}, {Aristidi},
  {Baffa}, {Behrend}, {Bl{\"o}cker}, {Bonhomme}, {Busoni}, {Cassaing},
  {Clausse}, {Colin}, {Connot}, {Delboulb{\'e}}, {Domiciano de Souza},
  {Driebe}, {Feautrier}, {Ferruzzi}, {Forveille}, {Fossat}, {Foy},
  {Fraix-Burnet}, {Gallardo}, {Giani}, {Gil}, {Glentzlin}, {Heiden},
  {Heininger}, {Hernandez Utrera}, {Hofmann}, {Kamm}, {Kiekebusch}, {Kraus},
  {Le Contel}, {Le Contel}, {Lesourd}, {Lopez}, {Lopez}, {Magnard}, {Marconi},
  {Mars}, {Martinot-Lagarde}, {Mathias}, {M{\`e}ge}, {Monin}, {Mouillet},
  {Mourard}, {Nussbaum}, {Ohnaka}, {Pacheco}, {Perrier}, {Rabbia}, {Rebattu},
  {Reynaud}, {Richichi}, {Robini}, {Sacchettini}, {Schertl}, {Sch{\"o}ller},
  {Solscheid}, {Spang}, {Stee}, {Stefanini}, {Tallon}, {Tallon-Bosc}, {Tasso},
  {Testi}, {Vakili}, {von der L{\"u}he}, {Valtier}, {Vannier}, \&
  {Ventura}}]{Petrov2007}
{Petrov}, R.~G., {Malbet}, F., {Weigelt}, G., {et~al.} 2007, \aap, 464, 1

\bibitem[{{Pinte} {et~al.}(2008){Pinte}, {M{\'e}nard}, {Berger}, {Benisty}, \&
  {Malbet}}]{Pinte2008}
{Pinte}, C., {M{\'e}nard}, F., {Berger}, J.~P., {Benisty}, M., \& {Malbet}, F.
  2008, \apjl, 673, L63

\bibitem[{{Pottasch} \& {Parthasarathy}(1988)}]{Pottasch1988}
{Pottasch}, S.~R. \& {Parthasarathy}, M. 1988, \aap, 192, 182

\bibitem[{{Preibisch} {et~al.}(2002){Preibisch}, {Brown}, {Bridges},
  {Guenther}, \& {Zinnecker}}]{Preibisch2002}
{Preibisch}, T., {Brown}, A.~G.~A., {Bridges}, T., {Guenther}, E., \&
  {Zinnecker}, H. 2002, \aj, 124, 404

\bibitem[{{Sylvester} {et~al.}(1996){Sylvester}, {Skinner}, {Barlow}, \&
  {Mannings}}]{Sylvester1996}
{Sylvester}, R.~J., {Skinner}, C.~J., {Barlow}, M.~J., \& {Mannings}, V. 1996,
  \mnras, 279, 915

\bibitem[{{Tatulli} {et~al.}(2011){Tatulli}, {Benisty}, {M{\'e}nard},
  {Varni{\`e}re}, {Martin-Za{\"i}di}, {Thi}, {Pinte}, {Massi}, {Weigelt},
  {Hofmann}, \& {Petrov}}]{Tatulli2011}
{Tatulli}, E., {Benisty}, M., {M{\'e}nard}, F., {et~al.} 2011, \aap, 531, A1+

\bibitem[{{Tatulli} {et~al.}(2007){Tatulli}, {Millour}, {Chelli}, {Duvert},
  {Acke}, {Hernandez Utrera}, {Hofmann}, {Kraus}, {Malbet}, {M{\`e}ge},
  {Petrov}, {Vannier}, {Zins}, {Antonelli}, {Beckmann}, {Bresson}, {Dugu{\'e}},
  {Gennari}, {Gl{\"u}ck}, {Kern}, {Lagarde}, {Le Coarer}, {Lisi}, {Perraut},
  {Puget}, {Rantakyr{\"o}}, {Robbe-Dubois}, {Roussel}, {Weigelt}, {Accardo},
  {Agabi}, {Altariba}, {Arezki}, {Aristidi}, {Baffa}, {Behrend}, {Bl{\"o}cker},
  {Bonhomme}, {Busoni}, {Cassaing}, {Clausse}, {Colin}, {Connot},
  {Delboulb{\'e}}, {Domiciano de Souza}, {Driebe}, {Feautrier}, {Ferruzzi},
  {Forveille}, {Fossat}, {Foy}, {Fraix-Burnet}, {Gallardo}, {Giani}, {Gil},
  {Glentzlin}, {Heiden}, {Heininger}, {Kamm}, {Kiekebusch}, {Le Contel}, {Le
  Contel}, {Lesourd}, {Lopez}, {Lopez}, {Magnard}, {Marconi}, {Mars},
  {Martinot-Lagarde}, {Mathias}, {Monin}, {Mouillet}, {Mourard}, {Nussbaum},
  {Ohnaka}, {Pacheco}, {Perrier}, {Rabbia}, {Rebattu}, {Reynaud}, {Richichi},
  {Robini}, {Sacchettini}, {Schertl}, {Sch{\"o}ller}, {Solscheid}, {Spang},
  {Stee}, {Stefanini}, {Tallon}, {Tallon-Bosc}, {Tasso}, {Testi}, {Vakili},
  {von der L{\"u}he}, {Valtier}, \& {Ventura}}]{Tatulli2007}
{Tatulli}, E., {Millour}, F., {Chelli}, A., {et~al.} 2007, \aap, 464, 29

\bibitem[{{Th\'e} {et~al.}(1994){Th\'e}, {de Winter}, \& {Perez}}]{The1994}
{Th\'e}, P.~S., {de Winter}, D., \& {Perez}, M.~R. 1994, \aaps, 104, 315

\bibitem[{{van Boekel} {et~al.}(2005){van Boekel}, {Dullemond}, \&
  {Dominik}}]{vanBoekel2005}
{van Boekel}, R., {Dullemond}, C.~P., \& {Dominik}, C. 2005, \aap, 441, 563

\end{thebibliography}

\appendix

\section{Searching for the best-fit models}

\begin{table*}[b]
\caption{The scanned parameter space in the two-component disk {\TGM}ing.
{\bf Upper part:} one-component disk model.
{\bf Lower part:} two-component disk model.
} 
\label{tab:TGM_range_1disk}
\label{tab:TGM_range_2disk}
\center

\setlength{\tabcolsep}{8pt}
\renewcommand{\arraystretch}{1.4}
\begin{tabular}{ccccccccccccccc}
\hline
\hline
Step                    &   $i(\degree)$      &  $\theta_{D}(\degree)$
& $T_{\mathrm{1,in}}($K$)$&   $q_1$           &$r_{\mathrm{1,in}}(\AU)$ & $\Delta r_1(\AU)$
& $T_{\mathrm{2,in}}($K$)$&   $q_2$           &$r_{\mathrm{2,in}}(\AU)$ & $\Delta r_2(\AU)$
                        \\
\hline
 1                      &  $ 0\enDash60$      &  $-90\enDash90$
& $800\enDash3000$      &  $0\enDash3$        & $0.05\enDash0.5$      & $0.001\enDash5$
\\
\hline
\hline
1                       &  $ 0\enDash60$      &  $-90\enDash90$
& $800\enDash3000$      &  $0\enDash3$        & $0.05\enDash0.5$      & $0.001\enDash5$
& $100\enDash 700$      &  $0\enDash3$        & $0.2 \enDash10 $      & $0.05 \enDash50$
\\
2                       &  $ 0\enDash50$      &  $-90\enDash90$
& $1000\enDash2500$     &  $0\enDash3$        & $0.15\enDash0.4$      & $0.003\enDash0.1$
& $250\enDash 500$      &  $0.2\enDash1.8$    & $0.4 \enDash 3 $      & $2.0  \enDash50$
\\
3                       &  $ 0\enDash40$      &  $-90\enDash90$
& $1200\enDash2000$     &  $0\enDash3$        & $0.15\enDash0.3$      & $0.008\enDash0.04$
& $350\enDash 450$      &  $0.5\enDash1.3$    & $0.6 \enDash1.2$      & $4.0  \enDash15$
\\
4                       &  $ 0\enDash35$      &  $-90\enDash90$
& $1300\enDash1900$     &  $0\enDash3$        & $0.18\enDash0.3 $     & $0.010\enDash0.03$
& $360\enDash 440$      &  $0.6\enDash1.1$    & $0.7 \enDash1.2$      & $5.0  \enDash12$
\\
5                       &  $ 0\enDash30$      &  $-90\enDash90$
& $1400\enDash1800$     &  $0.5      $        & $0.18\enDash0.25$     & $0.014\enDash0.024$
& $360\enDash 420$      &  $0.7\enDash1.0$    & $0.8 \enDash1.1$      & $6.0  \enDash10$
\\
6                       &  $ 0\enDash30$      &  $-90\enDash90$
& $1500\enDash1700$     &  $0.5      $        & $0.2 \enDash0.23$     & $0.016\enDash0.022$
& $370\enDash 410$      &  $0.75\enDash0.9$   & $0.85\enDash1.0$      & $7.0  \enDash9 $
\\
\hline
\end{tabular}
\end{table*}

\newcommand{\allTGMparaTwo}{\left(
i, \theta_D
,T_{\mathrm{1,in}},q_1,r_{\mathrm{1,in}},\Delta r_1
,T_{\mathrm{2,in}},q_2,r_{\mathrm{2,in}},\Delta r_2
,k_{\mathrm{halo}}
\right)}
\subsection{General description of the searching method}
We have searched for the best-fit one-component and two-component models
describing the visibility and SED data.
We established a grid in the multi-dimensional parameter space, and evaluate the  $\chi^2$ on each grid point to find the $\chi^2$-minimum.
A large parameter range was first scanned to roughly locate the global minimum.
Then, in further processing, we computed narrower grids around the global minimum to confine the parameters with higher precision.
In each step, we computed the models for all combinations of all parameters.

We treated the parameter $k_\mathrm{halo}$ in the following way.
For each combination of all the other parameters,
the modeled visibilities depend linearly on the halo fraction $k_\mathrm{halo}/(1+k_\mathrm{halo})$,
while the modeled fluxes are independent of $k_\mathrm{halo}$.
Therefore, the best-fit value of $k_\mathrm{halo}$ can be found with linear regression.

\subsection {Searching for the best-fit one-component disk model}
We searched for the best-fit one-component disk model
within the wide parameter ranges listed in Table \ref{tab:TGM_range_1disk}.
We divided the range of each parameter into 20 grid points
and computed $20^6\approx6\times10^7$ models for all combinations of all parameters.
No model with reasonable fitting could be found.

\subsection {Searching for the best-fit two-component disk model}
\begin{figure}
\includegraphics[scale=0.3,angle=-90]{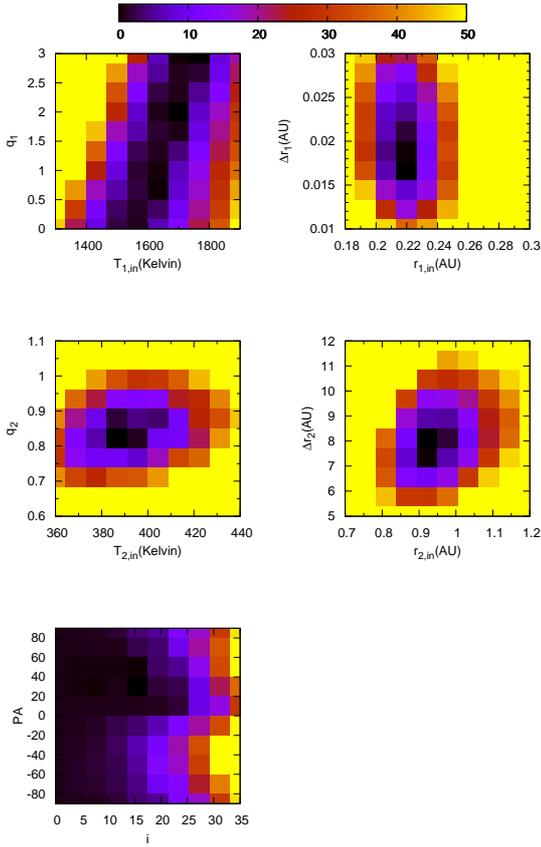}
\caption{
  $\chi^2$ maps showing
  $\Delta\chi^2=\chi^2-\chi^2_\mathrm{min}$ as function of the parameters in the two-componet disk model
  (results of the searching step 4).
  For each subset of parameters,
  the $\Delta\chi^2$ shown is the lowest value of all combinations of other parameters.
  }
\label{fig:chi2_2disk}
\end{figure}

In searching for the best-fit two-component disk model,
we start from wide parameter ranges and gradually zoom in (see the ranges listed in Table \ref{tab:TGM_range_2disk}).
The best-fit parameters are listed in Table \ref{tab:TGM_best}.
Due to the small radial thickness of the inner ring,
fits of equal quality (similar $\chi^2$ values) can be found for each $q_1$ value (see the $\chi^2$ map in Fig. \ref{fig:chi2_2disk}).
Therefore, we set $q_1=0.5$ 
\citep[corresponding to a flared irradiated disk;][]{Kenyon1987}
in the last two searching steps.

In each scanning step, we divided the range of each parameter into 10 grid points (except $q_1$ in the last two steps)
and computed the models for all combinations of all parameters.
In total we computed $4.2\times10^{10}$ models.

\end{document}